\documentclass[12pt]{elsart}

\usepackage{amssymb}
\usepackage{graphics}
\usepackage{graphicx}
\newcommand{\xx}{\mbox{\boldmath$x$}}
\newcommand{\yy}{\mbox{\boldmath$y$}}
\newcommand{\vv}{\mbox{\boldmath$v$}}
\newcommand{\pp}{\mbox{\boldmath$p$}}
\newcommand{\drho}{\delta\rho}
\newcommand{\dPhi}{\delta\Phi}
\begin{document}

\begin{frontmatter}


\title{\large Gravothermal Catastrophe and Tsallis' Generalized Entropy of 
Self-Gravitating Systems}
\author[taruya]{Atsushi Taruya}
\address[taruya]{Research Center for the Early Universe(RESCEU), 
School of Science, University of Tokyo, Tokyo 113-0033, Japan}
\ead{ataruya@utap.phys.s.u-tokyo.ac.jp}


\author[sakagami]{Masa-aki Sakagami}
\address[sakagami]{Department of Fundamental Sciences, FIHS, 
Kyoto University, Kyoto 606-8501, Japan}
\ead{sakagami@phys.h.kyoto-u.ac.jp}
\begin{abstract}
We present a first physical application of Tsallis' generalized entropy 
to the thermodynamics of self-gravitating systems.  
The stellar system confined in a spherical cavity of radius $r_e$ exhibits 
an instability, so-called gravothermal catastrophe,   
which has been originally investigated by Antonov (Vest.Leningrad 
Gros.Univ. 7 (1962) 135) and Lynden-Bell 
\& Wood (Mon.Not.R.Astron.Soc. 138 (1968) 495) on the basis of the 
maximum entropy principle for the phase-space distribution function. 
In contrast to previous analyses using the Boltzmann-Gibbs entropy, 
we apply the Tsallis-type generalized entropy to seek the equilibrium 
criteria. Then the distribution function of Vlassov-Poisson system 
can be reduced to the stellar polytrope system.  
Evaluating the second variation of Tsallis entropy and  
solving the zero eigenvalue problem explicitly, we find that the 
gravothermal instability appears in cases with polytrope index 
$n>5$. The critical point characterizing the onset of 
instability are obtained, which exactly matches with the results 
derived from the standard turning-point analysis. 
The results give an important suggestion that the Tsallis' 
generalized entropy is indeed applicable and viable 
to the long-range nature of the self-gravitating system. 
\end{abstract}
\end{frontmatter}
%
%
%
%
%
%
%
\section{Introduction}
\label{sec: intro}
%
%
%
In any subject of astrophysics and cosmology, many-body 
gravitating systems play an essential role. In general, dynamics of such 
systems are quite difficult to understand and the long-range nature of 
gravitational interaction prevents us from applying the statistical mechanics. 
Thermodynamics of self-gravitating systems also shows  
some peculiar features such as a negative specific heat and an absence of 
global entropy maxima, which greatly differ from usual thermodynamic systems.

To see the peculiarity of self-gravitating systems, 
consider a system confined within a spherical adiabatic wall. 
We assume that the particles in this system interact via Newton gravity 
and bounce elastically from the wall. Under keeping the energy and the 
total mass of the system constant, a thermodynamic description of 
self-gravitating 
systems leads to an interesting conclusion. When the central mass density 
is sufficiently high, no equilibrium state exists and the system 
can persistently undertake a strong central condensation. This instability 
is known as {\it gravothermal catastrophe}, originally investigated by 
Antonov\cite{Antonov1962} and Lynden-Bell and Wood\cite{LW1968} 
(see also Refs.\cite{HS1978,HNNS1978,Padmanabhan1989,Padmanabhan1990}). 
They define the entropy of distribution function in phase-space and 
seek the equilibrium condition whether the system exhibits the maximum
entropy state. In their analyses, they treat the Boltzmann-Gibbs entropy, 
\begin{equation}
  \label{eq: BG_entropy}
  S_{\rm BG}=-\int f\,\ln\,f\,\,d^3\xx d^3\pp, 
\end{equation}
where $f(\xx,\pp)$ is the distribution function in phase-space.

While the gravothermal catastrophe found in early studies 
has been widely accepted as a fundamental astrophysical process 
and plays an essential role in the 
dynamics of globular clusters\cite{BS1984,BT1987,EPI1987,MH1997}, 
a naive but natural question arises. 
Why should we use the Boltzmann-Gibbs entropy when looking for a probable 
entropy state ? The choice of entropy severely restricts the 
functional form of the distribution function $f$. In a spherically 
symmetric system with isotropic velocity distribution, the 
entropy (\ref{eq: BG_entropy}) leads to the isothermal gas 
distribution. The equilibrium configuration, however, cannot be 
unique in the self-gravitating systems. There remains a possibility 
of another choice of entropy in order to determine the most probable state.

Indeed, in the view of statistical mechanics and 
thermodynamics, it has recently been known that the standard 
formalism based on the Boltzmann-Gibbs entropy cannot deal with a variety 
of interesting physical problems and serious difficulties arise when 
applying the Boltzmann-Gibbs statistical mechanics (e.g, 
\cite{HD1994,SZF1995,K1997}). After introducing 
a family of generalized entropies by Tsallis \cite{T1988}, 
a new framework of thermodynamic structure has been extensively discussed   
\cite{CT1991,PP1997,TMP1998,MNPP2000,AMPP2001} 
and some physical applications has been presented successfully 
(e.g, \cite{B1996,LT1998,AA2001}; see \cite{T1999,AO2001} for 
comprehensive reviews). Although there still remains 
a fundamental issue on the consistency with the thermodynamic relations 
\cite{AMPP2001}, the new formalism is expected to be applied in 
a quite wide area including physics, astronomy, biology, economics, etc. 
Especially, in many astrophysical problems involving the 
long-range nature of gravity, the nonextensive generalization of 
statistical and/or thermodynamic treatment 
should deserve further consideration. At present, however, 
no specific examples have been reported theoretically, 
except for the experimental result \cite{HH2001} or observational 
evidence \cite{LKRQT1998}.

If one applies Tsallis' generalized entropy to the above problem in 
self-gravitating systems, the equilibrium condition is explored by using 
\begin{equation}
  \label{eq: TS_entropy}
  S_{q}=-\frac{1}{q-1}\int\, (f^{q}-f)\,\,d^3\xx d^3\pp, 
\end{equation}
instead of (\ref{eq: BG_entropy}). Here, the parameter $q$ is chosen 
as $q\neq1$, as a possible generalization of the Boltzmann-Gibbs entropy. 
Using (\ref{eq: TS_entropy}), an attempt to determine the most 
probable state has been made by Plastino and Plastino 
\cite{PP1993,PP1999}. They found that the equilibrium 
configuration reduces to the polytropic gaseous system, which has
been widely utilized in a study of the stellar structure 
\cite{Chandra1939,KW1990}.

Based on their result, 
we then pursue to investigate the stability 
of self-gravitating system confined within a box and try to answer the   
crucial question; how the equilibrium condition is altered when 
applying the entropy (\ref{eq: TS_entropy}) ?  
More basically, does the Tsallis type entropy correctly predict 
the equilibrium condition of a stellar polytrope system ?

This paper is organized as follows. In section 
\ref{sec: tsallis_entropy}, we briefly review that the most probable state 
determined by the Tsallis entropy reduces to the stellar polytrope. 
Then we proceed to the stability analysis based on 
the standard turning-point analysis in section \ref{sec: poincare}. 
In section \ref{sec: 2nd_variation}, stability/instability  
criterion is re-derived from the second variation of entropy.  
The stability/instability criterion can be obtained by solving zero-eigenvalue 
problem, which exactly matches the 
criterion from the standard turning-point analysis.  
Also, the geometrical construction of stability/instability criterion 
is discussed in terms of homology invariants. 
Finally, section 
\ref{sec: conclusion} is devoted to conclusions and discussion. 
%
%
%
\section{Tsallis entropy and stellar polytrope} 
\label{sec: tsallis_entropy}
%
%
%
%
In this section, for the notational convenience and the subsequent 
analyses, we start to review the work by Plastino \& Plastino \cite{PP1993} 
that the extremum state of the Tsallis entropy 
just reduces to the stellar polytrope in section \ref{subsec: Max_Ent}.  
Adopting this extremum state, 
the equilibrium configuration is then determined by solving the 
conditions of hydrostatic equilibrium in section 
\ref{subsec: Emden}. A family of equilibrium 
sequence referred to as the Emden solutions is obtained 
and characterized by the homology invariants, which is subsequently 
used in the stability analysis. 
%
%
%
%
\subsection{The principle of maximum entropy}
\label{subsec: Max_Ent}

Consider a system containing $N$ particles which are 
confined within a cavity of hard sphere. 
The radius of cavity is $r_e$ and each particle is 
assumed to have the same mass $m$. Then, 
all the information of this system can be described by a $N$-body 
distribution function $f_N(\xx_1,\cdots,\xx_N;\pp_1,\cdots,\pp_N;t)$, 
defined in the $6N$-dimensional phase-space. The evolution of such a system 
is governed by the collisionless Boltzmann equation, which is generally 
intractable. Instead of using the full distribution, 
it is better to employ a coarse-grained distribution function 
whose value at any phase-space point $(\xx,\pp)$ is averaged over 
some specified volume centered on $(\xx,\pp)$. 
In this treatment, the correlation between particles is smeared out. 
Thus, the system can be regarded as a gravitating gaseous system. 
The phase space distribution is simply described by the 
``one-particle distribution function'' $f(\xx,\pp;t)$, which 
greatly reduces the problem to a tractable level.

Adopting the mean-field treatment, we now investigate 
the equilibrium configuration and stability of a gravitating gaseous 
system with a help of statistical mechanics. In this approach, 
the equilibrium configuration can be 
determined by the principle of maximum entropy, which 
specifies the distribution function $f$ that maximizes the 
entropy. As is well-known, however, the global maximum does not exist in 
self-gravitating systems\cite{Antonov1962,LW1968}\cite{BT1987}. 
Hence, we try to look for the local extrema 
$\delta S=0$ and seek the criterion whether these extrema are really local 
maxima or not. 
Previously, most previous work has extensively discussed this issue on the
basis of the Boltzmann-Gibbs entropy (\ref{eq: BG_entropy}), in which case
the extremum solution reduces to the isothermal gas system. 
In this paper, with a particular attention to the Tsallis-type generalized 
entropy (\ref{eq: TS_entropy}), we seek the equilibrium configuration under 
the mass and the energy conservation: 
\begin{equation}
  \label{eq: mass}
  M\equiv\int\, f\,\,d^3\xx d^3\vv,   
\end{equation}
%
%
\begin{equation}
  \label{eq: energy}
  E= K+U \equiv 
  \frac{1}{2}\int\, \\v^{2}\,f\,\,d^3\xx d^3\vv\,\, 
  +\,\,\frac{1}{2}\int\,\Phi(\xx)\,f\,\,d^3\xx d^3\vv, 
\end{equation}
where the quantity $\Phi$ denotes gravitational potential: 
\begin{equation}
  \label{eq: potential}
  \Phi(\xx)=-G\,\int\,\,\frac{f(\yy,\vv)}{|\xx-\yy|}\,\,d^3\yy d^3\vv,    
\end{equation}
with $\vv=\pp/m$. Here, we adopted the standard 
definition of mean value for the mass $M$ and the energy $E$. The 
remark on the use of other definitions such as $q$-generalized mean 
value \cite{TMP1998} is discussed in section \ref{sec: conclusion}.

The extremum entropy state can be derived by varying $S_q$ with 
respect to $f$. Using the Lagrange multipliers $\alpha$ and $\beta$, 
the extremum solution subject to the constraints (\ref{eq: mass}) and
(\ref{eq: energy}) is obtained from 
\begin{eqnarray}
\delta S_q-\alpha\,\delta M-\beta\,\delta E=0,
\label{eq: 1st_variation}
\end{eqnarray}
which leads to 
\begin{eqnarray}
\int \,
\left\{-\frac{1}{q-1}(qf^{q-1}-1)-\alpha-\beta\left(\frac{1}{2}v^{2}+\Phi
\right)\right\} \delta f\,\,d^3\xx d^3\vv=0. 
\end{eqnarray}
Here the relation 
$\int \delta\Phi\, f\,d^3\xx d^3\vv=\int \Phi\, \delta f\,d^3\xx d^3\vv$ 
is used in deriving the above expression. Since the constraint 
(\ref{eq: 1st_variation}) must be satisfied independently of the 
choice of $\delta f$, we obtain 
\begin{equation}
\frac{1}{q-1}(q f^{q-1}-1)+\alpha+\beta\left(\frac{1}{2}v^{2}+\Phi
\right)=0,  
\end{equation}
which reduces to the following distribution function: 
\begin{equation}
  \label{eq: df_1}
  f(\xx,\vv)\,\, =\,\,A\,\,
\left[\Phi_0-\Phi(\xx)-\frac{1}{2}v^2\right]^{1/(q-1)}, 
\end{equation}
where we define the constants $A$ and $\Phi_0$:   
\begin{equation}
 A=\left\{\left(\frac{q-1}{q}\right)\beta\right\}^{1/(q-1)}, 
\,\,\,\,\,\,\,\,\,\,\,\,
\Phi_0= \frac{1-(q-1)\alpha}{(q-1)\beta}.
\end{equation}

The functional form of the distribution function (\ref{eq: df_1}) 
implies that the extremum solution is indeed equivalent to 
the polytropic gaseous systems \cite{BT1987}\cite{PP1993}.  
The density profile $\rho(r)$ at radius $r=|\xx|$ can be expressed 
using (\ref{eq: df_1}):  
\begin{eqnarray}
 \rho(\xx)\, &\equiv& \,\,\int \, f\,d^3\vv, 
\nonumber \\
 &=& \,\,4\sqrt{2}\pi\,\,B\left(\frac{3}{2},\,\frac{q}{q-1}\right)\,\,A\,
   \left\{\Phi_0-\Phi(\xx)\right\}^{\frac{1}{q-1}+\frac{3}{2}}
   \label{eq: def_rho}
\end{eqnarray}
with $B(a,b)$ being the $\beta$ function. On the other hand, 
in the case of isotropic velocity distribution,  
the pressure becomes
\begin{eqnarray}
 P(\xx)\, &\equiv& \,\,\int \,\frac{1}{3}v^2\,\,f\,\,d^3\vv, 
\nonumber\\
&=&\,\,\left(\frac{1}{q-1}+\frac{5}{2}\right)^{-1}\,\, \rho(\xx)\,\,
\left\{\Phi_0-\Phi(\xx)\right\}.
   \label{eq: def_pressure}
\end{eqnarray}
Thus, these two equations lead to the relation 
\begin{equation}
 P\,\,\propto\,\,\,\rho^{(5q-3)/(3q-1)}\,\,,  
\end{equation}
which corresponds to the polytropic equation of state, 
$P\propto\rho^{1+1/n}$, and the polytrope index $n$ is 
connected with Tsallis' $q$-parameter as follows:  
\begin{equation}
 n = \frac{1}{q-1}+\frac{3}{2}.
\label{eq: n_q}
\end{equation}

Note that the polytrope index $n$ should be positive in any 
astrophysical system. It has also been argued that the values 
of $n$ in the interval $0<n<3/2$ are unphysical\cite{BT1987}, 
which restricts the parameter $q$ to be larger than unity. 
Further, a simple argument shows that the locally maximum 
entropy can only be attained in the spherically symmetric configuration 
\cite{Antonov1962,LW1968}. Although the extremum solution (\ref{eq: df_1}) 
does not restrict any symmetry and any value of the parameter $q$,  
we hereafter restrict 
our attention to the spherically symmetric case with $q\geq1$.

The resultant distribution (\ref{eq: df_1}) and the relation (\ref{eq: n_q}) 
further imply that the polytrope gas is equivalent 
to the isothermal gaseous sphere in the limit $q\to1$ or $n\to +\infty$,  
which is in fact obtained in  adopting the Boltzmann-Gibbs entropy 
(\ref{eq: BG_entropy}). 
To check this consistency, let us take the limit $q\to1$. 
Introducing the new constant $T$: 
\begin{equation}
T\equiv(q-1) \left\{4\sqrt{2}\pi B\left(\frac{3}{2},\frac{q}{q-1}\right) 
\,A \right\}^{-(2q-2)/(3q-1)}, 
\label{eq: def_T}
\end{equation}
one can rewrite the distribution function (\ref{eq: df_1}) using the 
relation (\ref{eq: n_q}) and (\ref{eq: def_T}) as 
\begin{eqnarray}
f(r,v)&=&\frac{1} {4\sqrt{2}\pi B(3/2,n-1/2)} 
\nonumber \\
&&\times 
\frac{\rho(r)\,\,}{\left\{(n-3/2)\,T\,\rho^{1/n}(r)\right\}^{3/2}}\,\left[1
-\frac{v^2/2}{(n-3/2)\,T\,\rho^{1/n}(r)}\right]^{n-3/2}. 
\label{eq: DF_Tsallis}
\end{eqnarray}
Also the polytropic equation of state becomes
\begin{equation}
 P(r) =\left(\frac{n-3/2}{n+1}\right)\,T\, \rho^{1+1/n}(r).
\label{eq: pressure}
\end{equation}
Then using the fact that $B(3/2,n-1/2)\to(\pi/4n^3)^{1/2}$ 
in the limit $n\to \infty$, the distribution function asymptotically 
approaches 
\begin{eqnarray}
f(r,v) \stackrel{n\to+\infty}{\longrightarrow}
\frac{1}{(2\pi T)^{3/2}}\,\,\rho(r) e^{-v^2/2T}, 
\label{eq: f_isotherm}
\end{eqnarray}
and the equation of state reduces to that of the isothermal gas, 
$P=\rho T$. The expression (\ref{eq: f_isotherm}) shows that 
the velocity distribution is indeed Maxwellian and functional form 
of the distribution is fully specified by the 'temperature', 
$T$ (velocity dispersion). 
%
%
%
%
%
%
%
%
%
%
%
\subsection{Emden solution}
\label{subsec: Emden}

While the extremum state is determined by the variation of entropy,  
the distribution function (\ref{eq: df_1}) 
does not yet completely specify the equilibrium configuration. 
In the expression (\ref{eq: df_1}), 
the gravitational potential $\Phi$ appears, which implicitly 
involves $f$ again 
(see eq.(\ref{eq: potential})). To break this roundabout, 
an explicit expression for $\Phi$ or $\rho$ is needed.

In the spherically symmetric configuration, 
the gravitational potential given by (\ref{eq: potential}) satisfies 
the following Poisson equation:  
\begin{equation}
\label{eq: poisson_eq}  
 \frac{1}{r^2}\frac{d}{dr}\left(r^2\frac{d\Phi(r)}{dr}\right)=
  4\pi G\,\rho(r).
\end{equation}
Combining (\ref{eq: poisson_eq}) with (\ref{eq: def_rho}), we
obtain the ordinary differential equation for $\Phi$. 
Alternatively, a set of equations are derived using 
(\ref{eq: def_rho}), (\ref{eq: def_pressure}) and (\ref{eq: poisson_eq}):  
\begin{eqnarray}
& \frac{dP(r)}{dr}\,=&\,\,-\frac{Gm(r)}{r^2}\,\rho(r), 
\label{eq: hydro_1}
\\
& \frac{dm(r)}{dr}\,=&\,\,4\pi\rho(r)\,r^2, 
\label{eq: hydro_2}
\end{eqnarray}
which represent the hydrostatic equilibrium. 
The quantity $m(r)$ denotes the mass evaluated at the radius $r$
inside the wall. We then introduce the dimensionless quantities: 
\begin{equation}
\label{eq: dimensionless}
 \rho=\rho_c\,\left[\theta(\xi)\right]^n,\,\,\,\,\,\,
r=\left\{\frac{(n+1)P_{\rm c}}{4\pi G\rho_c^2}\right\}^{1/2}\,\xi, 
\end{equation}
which yields the following ordinary differential equation: 
\begin{equation}
 \theta''+\frac{2}{\xi}\theta'+\theta^n=0,
\label{eq: Lane-emden_eq}
\end{equation}
where prime denotes the derivative with respect to $\xi$. 
The quantities $\rho_c$ and $P_c$ in (\ref{eq: dimensionless}) 
are the density and the pressure at $r=0$, respectively. To 
obtain the physically relevant solutions, 
we solve equation (\ref{eq: Lane-emden_eq}) with the following 
boundary condition:   
\begin{equation}
 \theta(0)=1, \,\,\,\,\,\,\,\theta'(0)=0.    
\label{eq: boundary}
\end{equation}
A family of solutions  
satisfying (\ref{eq: boundary}) is referred to as the {\it Emden
solution}, 
which is well-known in the subject of stellar structure and details of 
the solutions can be found in standard textbooks (e.g., see 
Chap.IV of ref.\cite{Chandra1939}).  
Except for the few cases $n=0$, $1$ and $5$, 
the Emden solution cannot be expressed in terms of the elementary
functions. So the solution with general index $n$ is obtained
numerically.

Notice the fact that equation (\ref{eq: Lane-emden_eq}) is invariant 
under the transformation $\xi \to A\xi$, $\theta \to
A^{-2/(n-1)}\theta$, where $A$ is an arbitrary constant. 
This implies that, given a solution with some 
value of $\theta(0)$, we can obtain the solution with any other value of  
$\theta(0)$ by simple rescaling. Therefore, only one of the two 
integration constants in (\ref{eq: Lane-emden_eq}) is actually 
non-trivial. This fact allows us to reduce the degree of equation from
two to one by suitable choice of variables.  One such set of
variables is 
\begin{eqnarray}
 u &\equiv& 
\frac{4\pi r^3\rho(r)}{m(r)}=-\frac{\xi\theta^n}{\theta'},
\label{eq: def_u}
\\
 v &\equiv& 
\frac{\rho(r)}{P(r)}\,\,\frac{Gm(r)}{r}
=-(n+1)\frac{\xi\theta'}{\theta}. 
\label{eq: def_v}
\end{eqnarray}
which are called {\it homology
invariants}\cite{Chandra1939,KW1990}. 
That is, for a fixed $n$, all the solutions depicted in $(u,v)$-plane 
lie on the same trajectory. In terms of these variables, the equation 
(\ref{eq: Lane-emden_eq}) can be written as  
\begin{equation}
 \label{eq: uv_eqn}
  \frac{u}{v}\,\frac{dv}{du}=\frac{(n+1)(u-1)+v}{(n+1)(3-u)-nv}.
\end{equation}

\begin{figure}
  \begin{center}
    \includegraphics*[width=10cm]{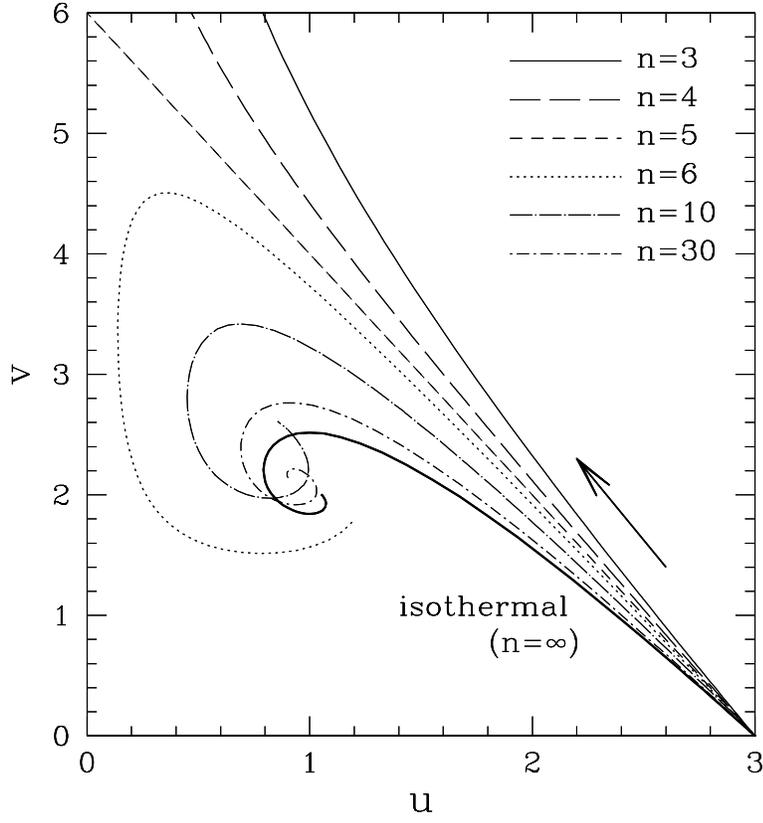}
  \end{center}
    \caption{Trajectories of Emden solution in $(u,v)$-plane}
    \label{fig: uv_plane1}
\end{figure}

Figure \ref{fig: uv_plane1} shows the Emden solutions  
for various polytrope indices. As indicated by 
the boundary condition, (\ref{eq: boundary}), 
all the trajectory start from $(u,v)=(3,0)$, which represents 
the center of the configuration, $r=0$.  
As the radius $r$ increases, the trajectory monotonically moves to 
the upper-left direction in $(u,v)$-plane, as marked by arrow.  
For larger radius, the solution with polytrope index $n>5$ 
spirals around a fixed point, while the trajectories with $n<5$ 
continue to approach the point $(u,v)=(0,\infty)$.   
The marginal case is $n=5$. The trajectory monotonically changes 
and asymptotically reaches $(u,v)=(0,6)$. 
These differences can be explained by the behavior of outer envelope
$\rho(r)$ as follows. 
In cases of $n<5$, the density $\rho(r)$ falls off rapidly 
and it eventually vanishes at a finite size, where the pressure also
becomes zero. Since the mass is finite and the polytropic relation 
$P\propto\rho^{1+1/n}$ holds, the quantities $(u,v)$ becomes
$(u,v)=(0,\infty)$ from (\ref{eq: def_u}) and (\ref{eq: def_v}). 
On the other hand, when $n>5$, the outer profile of the density
falls off less steeply than $\rho\propto r^{-5}$\cite{BT1987}. In this
case, the trajectory in $(u,v)$-plane cannot reach $u=0$. It must 
be bounded within the interval $0<u<3$ and shows the 
oscillatory behavior, as depicted in Figure \ref{fig: uv_plane1}.

The characteristic feature seen in Figure \ref{fig: uv_plane1} provides us 
an important suggestion about stability of the polytrope  
gas sphere. This point will be clarified by 
the turning-point analysis and
the second variation of entropy in the subsequent section.
%
%
%
%
\section{Stability/instability criterion from the turning-point analysis}
\label{sec: poincare}

We are specifically concerned with the stability/instability 
of static equilibria from the thermodynamic point of view.   
For this purpose, following the discussion in 
\cite{LW1968,Padmanabhan1989,Padmanabhan1990}, 
we first apply the standard-turning point analysis to the 
equilibrium sequence obtained in the previous section.

Recall that the system is confined within the spherical 
adiabatic wall of radius $r_e$. 
For fixing polytrope index $n$, 
the local extremum satisfying $\delta S_q=0$ is 
characterized by the energy $E$, mass $M $and radius of rigid 
sphere $r_e$. In the absence of instability,    
any value of $E(-\infty<E<\infty)$, $M(0<M<\infty)$ 
and $r_e(0<r_e<\infty)$ are accommodated by a suitable choice of 
$\rho_c$ and $P_c$, but, there exists a lower bound on 
the dimensionless quantity $Er_e/GM^2$ for the polytrope gas solutions. 
To show this, we compute the total energy contained within 
the hard sphere of a radius $r_e$. The kinetic and 
potential energies, $K$ and $U$ are respectively expressed as 
\begin{equation}
  \label{eq: K_energy}
  K= \frac{3}{2} \int_0^{r_e} dr\,4\pi r^2\, P(r),
\end{equation}
and 
\begin{eqnarray}
  \label{eq: U_energy} 
  U &=& -\int_0^{r_e}dr\,\frac{Gm}{r}\,\frac{dm}{dr} 
\,  = \, -\frac{G}{2} \int_0^{r_e}\,\frac{dr}{r} \frac{d}{dr}(m^2)
\,  = \, -\frac{GM^2}{2r_e} -\frac{1}{2}\int_0^{r_e}dr\, \frac{Gm^2}{r^2}
\nonumber \\
&=& \, 4\pi r_e^3 P_e \,- \,3 \int_0^{r_e} dr\,4\pi r^2\, P(r). 
\end{eqnarray}
Combining these results and using (\ref{eq: int_3}) in appendix A, 
the total energy $E$ can be written in a compact form:  
\begin{eqnarray}
  \label{eq: tot_energy}
  E &=& K+U=\,4\pi r_e^3 P_e -\frac{3}{2} \int_0^{r_e} dr\,4\pi r^2\, P(r)
\nonumber \\
  &=& \frac{1}{n-5} 
\left[  \frac{3}{2}\left\{\frac{GM^2}{r_e} -(n+1)\frac{MP_e}{\rho_e}\right\}
  +(n-2) 4\pi r_e^3 P_e \right],
\end{eqnarray}
where the subscript $(_{e})$ means a quantity evaluated at the 
boundary $r_{e}$. Then,  
the dimensionless quantity $\lambda$ is introduced and 
expressed in terms of homology invariants at the boundary $r=r_e$:
\begin{equation}
  \label{eq: ergm2}
   \lambda\equiv - \frac{Er_e}{GM^2} = - \frac{1}{n-5}\,
  \left[\frac{3}{2}\left\{1-(n+1)\frac{1}{v_e}\right\}+(n-2)
    \frac{u_e}{v_e}\right].  
\end{equation}

\begin{figure}
  \begin{center}
    \includegraphics*[width=15cm]{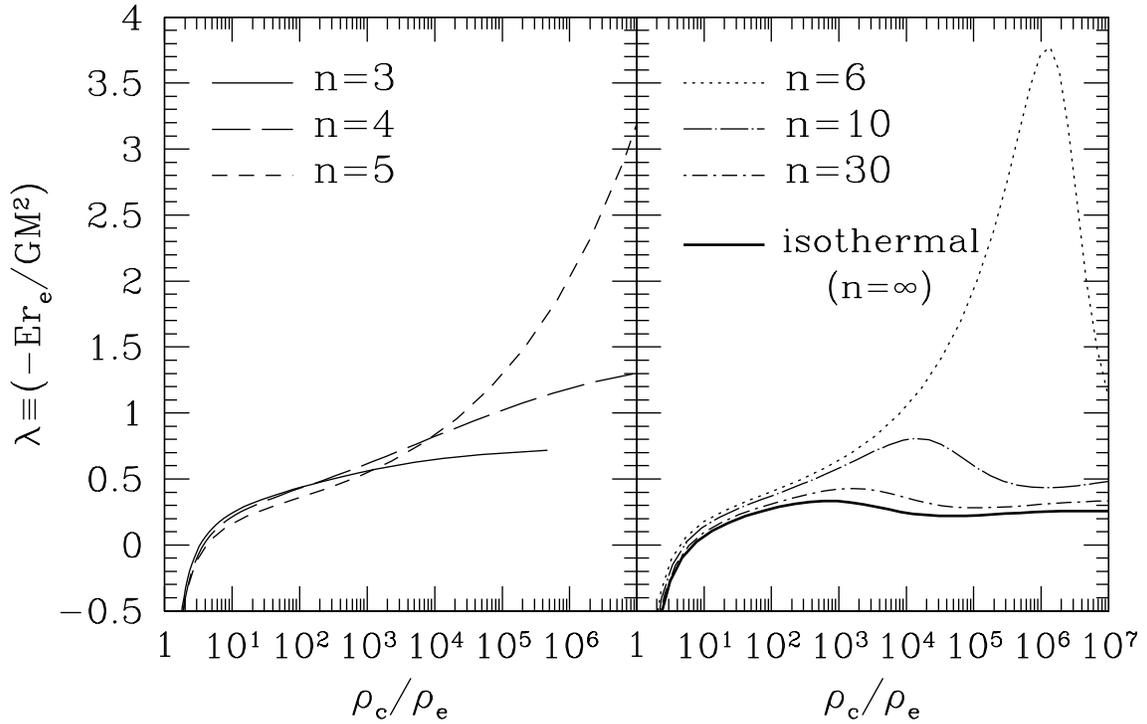}
  \end{center}
    \caption{Energy-radius-density contrast relationship}
    \label{fig: trajectory}
\end{figure}

Figure \ref{fig: trajectory} shows the quantity $\lambda$ 
as a function of the ratio of the central density to that at the 
boundary, $\rho_c/\rho_e$. In each panel, lines indicate 
a series of local extrema with a different polytrope index $n$. 
Note that all solutions satisfying $\delta S_q=0$ must lie on 
this curve.

At a first glance of Figure \ref{fig: trajectory}, we notice that
$\lambda$-curves are bounded from above and have peaks in 
the case of $n>5$ ({\it right panel}).
We call these peaks  critical points.
On the  other hand, curves for $n\leq5$ monotonically increase
({\it left panel}).   
It follows that, in the case of $n>5$, several extremum states 
correspond to a single value of $\lambda$ in some range of $\lambda$.   
Thus we can deduce the existence of
unstable state for $n>5$ as follows. 
Suppose that a polytropic gaseous sphere with small 
radius $r_e\to 0,~~(\rho_c/\rho_e\to1)$ is stable, i.e., $\delta^2S_q<0$
(see Appendix C for rigorous proof). 
Due to the continuity, entropy $S_q$ must be local maximum 
along each $\lambda$-curve. This is true as long as 
$\lambda$ increases monotonically. When the 
ratio $\rho_c/\rho_e$ exceeds the critical point, 
however, the extremum configuration cannot be stable. If this is stable, 
the entropy has to be a local maximum there, which implies that 
there exists a local minimum between these two local maxima. 
This contradicts with the fact that 
all extremum state must lie on the $\lambda$-curve. 
Therefore, local maximum of the entropy can only be attained 
at the density $\rho_c/\rho_e$ below the critical point and   
the stability/instability criterion is thus obtained from the 
condition that the quantity 
\begin{equation}
\frac{d\lambda}{d(\rho_c/\rho_e)}=0 
\label{eq: condition_a}
\end{equation}
first vanishes as increasing the ratio $\rho_c/\rho_e$. 
From the monotonicity of $\rho(r)$,  
this condition implies ${d\lambda}/{d\xi_e} =0$.  
Therefore, using the dimensionless quantities 
(\ref{eq: dimensionless}) and the equation (\ref{eq: Lane-emden_eq}),  
the stability/instability criterion reduces to 
\begin{equation}
0=\frac{d\lambda}{d\xi_e} \equiv 
\frac{n-2}{n-5}\,\, \frac{g(u_e,v_e)}{2v_e\,\xi_e}, 
\label{eq: stability_!!}
\end{equation}
where 
\begin{eqnarray}
&&g(u_e,v_e) \nonumber\\
&&~~=\,
4u_e^2+2u_ev_e-\left\{8+3\,\left(\frac{n+1}{n-2}\right)\right\}u_e
 -\frac{3}{n-2}\,v_e+3\,\left(\frac{n+1}{n-2}\right).
\label{eq: stability_g}
\end{eqnarray}
As will be shown in the next section, especially in eq.(\ref{eq: stability_!}),
this criterion can be derived also by means of an explicit method, i.e.,
evaluation of the second order variation of the entropy.

In Table \ref{tab: ErGM}, numerical values of dimensionless quantities 
$\lambda$ and $\rho_c/\rho_e $ evaluated at the critical point 
are summarized. Table \ref{tab: ErGM} shows that both two quantities 
decrease as $n$ increases. In the limit $n\to\infty$(or $q\to1$), they 
asymptotically approach the well-known results of the isothermal sphere 
\cite{Antonov1962,LW1968,HS1978}\cite{Padmanabhan1989}.  
%
%
%
%
%
%
%
%
%
%
%
\section{Stability criterion from the second variation of entropy}
\label{sec: 2nd_variation}
%
%
%
%

In this section, we re-consider the stability/instability  
of static self-gravitating system 
by evaluating the second variation 
of entropy. Then, the criterion (\ref{eq: stability_!!}) is 
re-derived, independently of the turning-point analysis.

According to the principle of maximum entropy, 
the equilibrium state can be attained only when the second variation of 
entropy $\delta^2S_q$ around the extremum solution becomes negative, 
$\delta^2S_q<0$. Conversely, the solution becomes unstable if one obtains 
$\delta^2S_q>0$.  The condition $\delta^2S_q=0$ corresponds to the 
neutral case, in which the equilibrium configuration becomes neither 
stable nor unstable. Thus, the stability/instability criterion can be 
extracted from $\delta^2S_q=0$.

Consider the variation of entropy around the extremum configuration, 
fixing the energy $E$ and the mass $M$. 
We deal with the density perturbation 
$\drho$ around the polytropic gaseous state $\rho(r)$. 
To be specific, we focus on the radial mode of the density 
perturbations under the condition:
\begin{equation}
 \int_0^{r_e}\, dr\,\, 4\pi r^2\drho(r)=0, 
\label{eq: del_mass}
\end{equation}
so as to satisfy the mass conservation, $\delta M=0$. 
In appendix B, variation of entropy around the equilibrium
configuration is computed up to the second order. The resultant expression 
for $\delta^2S_q$ then becomes 
\begin{eqnarray}
&& \delta^2S_q=\,\,-\frac{\tilde{B}_n}{T^{(3/2)/(n-3/2)}}\,\frac{n+1}{n-3/2}\,\,
  \left[\int d^3\xx \left\{\frac{\drho\dPhi}{2T}+\frac{n-3/2}{2n}
\frac{(\drho)^2}{\rho^{1-1/n}}\right\}
\right.
\nonumber \\
&& ~~~~+ \left. \frac{n(n+1)}{(n-3/2)^2}\,\,
\frac{1}{3T^2 W} 
\left\{\int d^3\xx\left(\Phi+\frac{3}{2}\,\frac{n-3/2}{n}\,T\,\rho^{1/n}\right)
\drho\right\}^2
\right],
\label{eq: 2nd_S} 
\end{eqnarray}
where the variable $\dPhi$ denotes the potential related to 
the perturbation $\drho$: 
\begin{equation}
 \frac{1}{r^2}\frac{d}{dr}\left(r^2\frac{d\,\dPhi}{dr}\right)
= 4\pi G \,\drho(r),
\label{eq: perturb_Poisson}
\end{equation}
and the quantities $\tilde{B}_n$ and $W$ are respectively given by  
\begin{eqnarray}
\tilde{B}_n &=& 
\frac{B(3/2,n+1/2)}{\left\{ B(3/2,n-1/2)\right\}^{(n-1/2)/(n-3/2)}}
\,\,\left[4\sqrt{2}\pi\,\left(n-\frac{3}{2}\right)^{3/2}\right]^{-1/(n-3/2)}, 
\nonumber \\
W &=& \int d^3\xx \,\rho^{1+1/n}, 
\nonumber
\end{eqnarray}
which reduce to $\tilde{B}_n\to1$ and $W\to M$, in the limit of 
$n\to\infty$(or $q\to1$).

Based on the result (\ref{eq: 2nd_S}), 
we now discuss the existence or the absence of perturbation mode 
satisfying $\delta^2S_q=0$ for given background 
$\rho(r)$. To simplify the analysis, it is convenient 
to introduce the new variable $Q(r)$ defined by 
\begin{equation}
 \drho(r)=\frac{1}{4\pi r^2}\,\frac{dQ(r)}{dr}. 
\end{equation}
Then the mass conservation (\ref{eq: del_mass}) implies the following boundary
condition: 
\begin{equation}
 Q(0)=Q(r_e)=0.
\label{eq: boundary_con_Q}
\end{equation}
In terms of $Q(r)$, equation (\ref{eq: 2nd_S}) becomes 
\begin{eqnarray}
&&\delta^2 S_q = \frac{\tilde{B}_n}{T^{(3/2)/(n-3/2)}}\,\frac{n+1}{n-3/2}
\,\,A[Q],
\end{eqnarray}
where 
\begin{eqnarray}
&& A[Q]\equiv - \int_0^{r_e} dr \left\{\frac{\dPhi}{2T}\frac{dQ}{dr}+
\frac{n-3/2}{n}\frac{1}{8\pi r^2
\rho^{1-1/n}}\left(\frac{dQ}{dr}\right)^2
\right\}
\nonumber \\
&&~~~~~-\frac{1}{3 T^2 W} \frac{n(n+1)}{(n-3/2)^2} 
\left[ \int_0^{r_e}dr \left\{\Phi+\frac{3}{2}\left(\frac{n-3/2}{n}\right)\,T\, 
\rho^{1/n}\right\} \frac{dQ}{dr}\right]^2.
\label{eq: A_term}
\end{eqnarray}
Since both $\tilde{B}_n$ and $T$ are positive, 
the condition $\delta^2S_q=0$ is translated to $A[Q]=0$. 
Thus, apart from the pre-factor, we pay attention to the term $A[Q]$. 
Integrating (\ref{eq: A_term}) by parts and 
using (\ref{eq: perturb_Poisson}), we obtain
\begin{eqnarray}
 A[Q]&=& \frac{1}{2}\int_0^{r_e} dr \,Q(r) 
\left\{ 
 \frac{G}{Tr^2}+\frac{n-3/2}{n}
\frac{d}{dr}\left(\frac{1}{4\pi r^2\,\rho^{1-1/n}}\frac{d}{dr}\right) 
\right\} Q(r) 
\nonumber \\
&-&
\frac{1}{3T^2W}\frac{n(n+1)}{(n-3/2)^2}\left[
\int_0^{r_e}dr\left\{\frac{d\Phi}{dr}+
\frac{3}{2}\left(\frac{n-3/2}{n^2}\right)\,T\,\rho^{-1+1/n}\frac{d\rho}
{dr}\right\}Q
\right]^2
\nonumber \\
&\equiv& 
-\int_0^{r_e} dr_1 \int_0^{r_e} dr_2 \,\, Q(r_1)\hat{K}(r_1,r_2)Q(r_2),
\label{eq: A_quad}
\end{eqnarray}
where the kernel $\hat{K}$ is given by 
\begin{eqnarray}
  \label{eq: kernel}
\hat{K}(r_1,r_2)\, &=& \,\, 
-\frac{1}{2}\,\delta_D(r_1-r_2)
\left\{\frac{n-3/2}{n}\frac{d}{dr_1}
\left(\frac{1}{4\pi r_1^2\,\left\{\rho(r_1)\right\}^{1-1/n}}
\frac{d}{dr_1}\right)+\frac{G}{Tr_1^2}\right\}
\nonumber \\
&&~~~~~~~ +\,\frac{1}{3T^2W}\,\frac{n(n+1)}{(n-3/2)^2}\, F(r_1)F(r_2) 
\end{eqnarray}
with quantity $F(r)$ being 
\begin{equation}
 F(r)= \frac{d\Phi}{dr}+\frac{3}{2}\left(\frac{n-3/2}{n^2}\right)\,T\,
\rho^{-1+1/n}\frac{d\rho}{dr}.
\end{equation}
Therefore, stability of the background configuration 
$\rho(r)$ is deduced from the following eigenvalue equation: 
\begin{equation}
 \int_0^{r_e} dr' \,\hat{K}(r,r') Q(r')=\kappa \,Q(r).
\label{eq: eigen_eq}
\end{equation}

From (\ref{eq: eigen_eq}), the stable condition $\delta^2 S_q<0$
indicates that minimum eigenvalue $\kappa_{\rm min}$ 
should be positive, while the local minimum of the entropy $\delta^2
S_q>0$ means $\kappa_{\rm min}<0$. Hence, the 
boundary between stable and the unstable configuration corresponds to 
the condition $\kappa_{\rm min}=0$. Notice that for a given polytrope 
index $n$, the minimum eigenvalue $\kappa_{\rm min}$ of the system with 
$(E,M)$ depends on the radius of the rigid sphere $r_e$.  For sufficiently 
small $r_e$, it can be shown that the configuration should be 
stable and we obtain the positive eigenvalue, $\kappa_{\rm min}>0$ (see 
appendix C). Therefore, to seek the boundary of the stability, 
it is sufficient to investigate the condition $\kappa_{\rm min}=0$ 
when increasing $r_e$. That is, the stability/instability criterion 
can be extracted from the following zero-eigenvalue equation:   
\begin{eqnarray}
\int_0^{r_e} dr' \,\hat{K}(r,r') Q(r')=0,
\nonumber
\end{eqnarray}
which yields 
\begin{eqnarray}
&&\hat{L}\,Q(r)\equiv \left[\frac{d}{dr}\left\{\frac{1}{4\pi r^2\rho}
\left(\frac{P}{\rho}\right)\frac{d}{dr}\right\}+\frac{n}{n+1}\frac{G}{r^2}\right]Q(r)
\nonumber \\
&& ~~~~~~~~~~~~
=\frac{2}{3}\,\frac{n-3/2}{n+1}\,\frac{1}{\displaystyle \int_0^{r_e}dr'\,4\pi r^2 P(r')}
\frac{Gm(r)}{r^2}\int_0^{r_e} dr' \frac{Gm(r')}{r'^2}Q(r').
\label{eq: zero_eigen_eq}
\end{eqnarray}
In deriving the above expression, we have utilized the relation 
(\ref{eq: pressure}) and the hydrostatic equilibrium conditions, 
(\ref{eq: hydro_1}) and (\ref{eq: hydro_2}).

The zero-eigenvalue equation (\ref{eq: zero_eigen_eq}) 
is integro-differential equation which seems 
intractable at a first glance, however, 
the solution satisfying boundary condition (\ref{eq: boundary_con_Q}) 
is luckily obtained from the knowledge of background configuration, 
$\rho(r)$ or $\Phi(r)$. 
To construct the solution $Q(r)$, first note the action of the operator 
$\hat{L}$ on $4\pi r^3 \rho$ and $m(r)$: 
\begin{eqnarray}
 \hat{L}\,(4\pi r^3 \rho) = \frac{d}{dr}\left\{-\frac{n}{n+1}\,\frac{Gm(r)}{r}+
  3\frac{P}{\rho}\right\} +\frac{n}{n+1} 4\pi G r\rho= 
\frac{n-3}{n+1}\,\frac{Gm(r)}{r^2},
\nonumber
\end{eqnarray}
and 
\begin{eqnarray}
\hat{L}\,m(r) = \frac{d}{dr}\left(\frac{P}{\rho}\right)+\frac{n}{n+1}
\frac{Gm(r)}{r^2} =\frac{n-1}{n+1}\,\frac{Gm(r)}{r^2},  
\nonumber
\end{eqnarray}
where we have used the following relations: 
\begin{equation}
\frac{P}{\rho}\,\frac{d\ln \rho}{d \ln r}= -\frac{n}{n+1}\frac{Gm}{r},
~~~~~~~~~~
\frac{d}{dr}\left(\frac{P}{\rho}\right)=-\frac{1}{n+1}\,\frac{Gm}{r^2}.  
\end{equation}
From the indication of these equations, we put the ansatz of the solution,  
\begin{equation}
 Q(r) = c_1\,4\pi r^3 \rho(r) +c_2\, m(r), 
\label{eq: ansatz_Q} 
\end{equation}
and determine the coefficients $c_1$ and $c_2$ by substituting 
(\ref{eq: ansatz_Q}) into (\ref{eq: zero_eigen_eq}). We then have 
\begin{equation}
 \frac{n-3}{n+1}\,\,c_1+\frac{n-1}{n+1}\,\,c_2 =\frac{2}{3}\,\,
\frac{n-3/2}{n+1}\,\,
\frac{\displaystyle \int_0^{r_e}dr'\,\frac{Gm(r')}{r'^2}Q(r')}
{\displaystyle \int_0^{r_e}dr'\,4\pi r'^2P(r')}\equiv \Lambda.
\label{eq: c1c2_a}
\end{equation}
Further, recall that the ansatz 
(\ref{eq: ansatz_Q}) must satisfy the boundary condition 
(\ref{eq: boundary_con_Q}). Since the condition $Q(0)=0$ is 
automatically satisfied, the remaining condition $Q(r_e)=0$ requires 
\begin{equation}
4\pi\, r_e^3\,\rho_e\,\,c_1+\,M\,c_2=0. 
\label{eq: c1c2_b}
\end{equation}
Equations (\ref{eq: c1c2_a}) and (\ref{eq: c1c2_b}) specify the coefficients, 
which can be expressed in terms of the homology invariants (see
definitions [\ref{eq: def_u}][\ref{eq: def_v}]):  
\begin{equation}
 c_1=\frac{(n+1)\Lambda}{n-3-(n-1)u_e},~~~~~~~~~~~ 
 c_2=-\,\frac{(n+1)\,u_e\,\Lambda}{n-3-(n-1)u_e}.
\label{eq: coeff}
\end{equation}

The solution (\ref{eq: ansatz_Q}) with (\ref{eq: coeff}) seems  
still uncertain because of the quantity $\Lambda$ in 
coefficients, which implicitly depends on the solution (\ref{eq:
ansatz_Q}) itself. This means that the value of the coefficients $c_1$ 
and $c_2$ should be further constrained so that the local expression 
(\ref{eq: ansatz_Q}) indeed satisfies the non-local equations. 
In other words, the consistency between (\ref{eq: ansatz_Q}) and 
$\Lambda$ puts the condition for the background quantities, 
$\rho_e$, $P_e$ and $M$ evaluated at $r=r_e$. This is just the 
stability/instability  
criterion we wish to clarify.

Now, substituting the solution (\ref{eq: ansatz_Q}), we evaluate the quantity
$\Lambda$ explicitly.  
\begin{equation}
\Lambda= \frac{2}{3}\,\frac{n-3/2}{n+1}\,\,
\frac{\displaystyle c_1\int_0^{r_e} dr\,\frac{Gm}{r^2}\,4\pi r^3 \rho + 
c_2 \int_0^{r_e}  dr\,\frac{G\,m^2}{r^2} 
}{\displaystyle \int_0^{r_e}dr \,4\pi r^2\,P}.
\label{eq: Lambda}
\end{equation}
The integrals in the right hand side of 
(\ref{eq: Lambda}) can be 
rewritten repeating the integration by parts as shown in 
Appendix A.
Together with the coefficients (\ref{eq: coeff}),  
substitution of (\ref{eq: int_1})-(\ref{eq: int_3}) into (\ref{eq:
 Lambda}) leads to the stability/instability criterion. 
In terms of $u$-$v$ variables, this gives
\begin{eqnarray}
 1 &=& \frac{2}{3}\,\frac{n-3/2}{n-3-(n-1)u_e}\,
\left[
\frac{(n-5)u_e}{2u_e+v_e-n-1}+3 
\right.
\nonumber\\
&&~~~~~~~~~~~~~~~~~~~~~~~~~~~~~~~~~~\left. 
-u_e
\left\{\frac{(n-5)\{2u_e+v_e\}}{2u_e+v_e-n-1}+6\right\} 
\right].
\nonumber 
\end{eqnarray}
After some algebra, the above equation finally reduces to the following 
quadratic form:  
\begin{eqnarray}
 0\,&=&\, 4u_e^2+2u_ev_e -
\left\{8+3\left(\frac{n+1}{n-2}\right)\right\}u_e-\frac{3}{n-2}\, v_e
+3\left(\frac{n+1}{n-2}\right)
\nonumber\\
&=&\, g(u_e,v_e),
\label{eq: stability_!}
\end{eqnarray}
which exactly matches the criterion derived from the turning-point analysis 
(see eqs.[\ref{eq: stability_!!}][\ref{eq: stability_g}]).

Equation (\ref{eq: stability_!}) is the main result of our analysis.      
It determines the critical point in the $(u,v)$-plane, where the 
local extremum entropy is neither maximum nor minimum 
\cite{Padmanabhan1989}.

To see the geometrical meaning of this criterion explicitly, 
we translate the result (\ref{eq: stability_!}) 
into the constraint in the $(u,v)$-plane. 
In the previous section, we see that a family of equilibrium sequences 
is characterized by the parameters, $r_e$, $E$ and $M$. 
These parameters specify the value of $\lambda$ defined 
in (\ref{eq: ergm2}). 
In other words, such a polytrope system must lie on the straight line: 
\begin{equation}
\label{eq: u-v_line}
 v\,=\,-\,\,\frac{n+1}{(n-5)\lambda-3/2}\,\,\left[\frac{n-2}{n-1}\,u
-\frac{3}{2}\right].
\end{equation}
from equation (\ref{eq: ergm2}). Note also that the polytrope solution must 
lie on the $u$-$v$ trajectory as shown in Figure \ref{fig: uv_plane1}.   
Hence, the equilibrium state with fixed parameter set $(r_e,E,M)$ 
can exist only if the $u$-$v$ curve intersects the straight 
line (\ref{eq: u-v_line}). From the configuration of the $u$-$v$ trajectory, 
we notice that the quantity $\lambda$ is bounded 
from above, $\lambda\leq\lambda_c$ in the case of $n>5$. 
That is, for the lines (\ref{eq: u-v_line}) with $\lambda$ larger than 
$\lambda_c$, the intersection ceases to exist. 
Since the critical value $\lambda_c$ must satisfy 
$d\lambda/d r=0$, this yields (\ref{eq: stability_!!}) or 
(\ref{eq: stability_!}) and the condition $g(u_e,v_e)=0$ determines  
the critical point along each $u$-$v$ trajectory. 

\begin{figure}
  \begin{center}
    \includegraphics*[width=15cm]{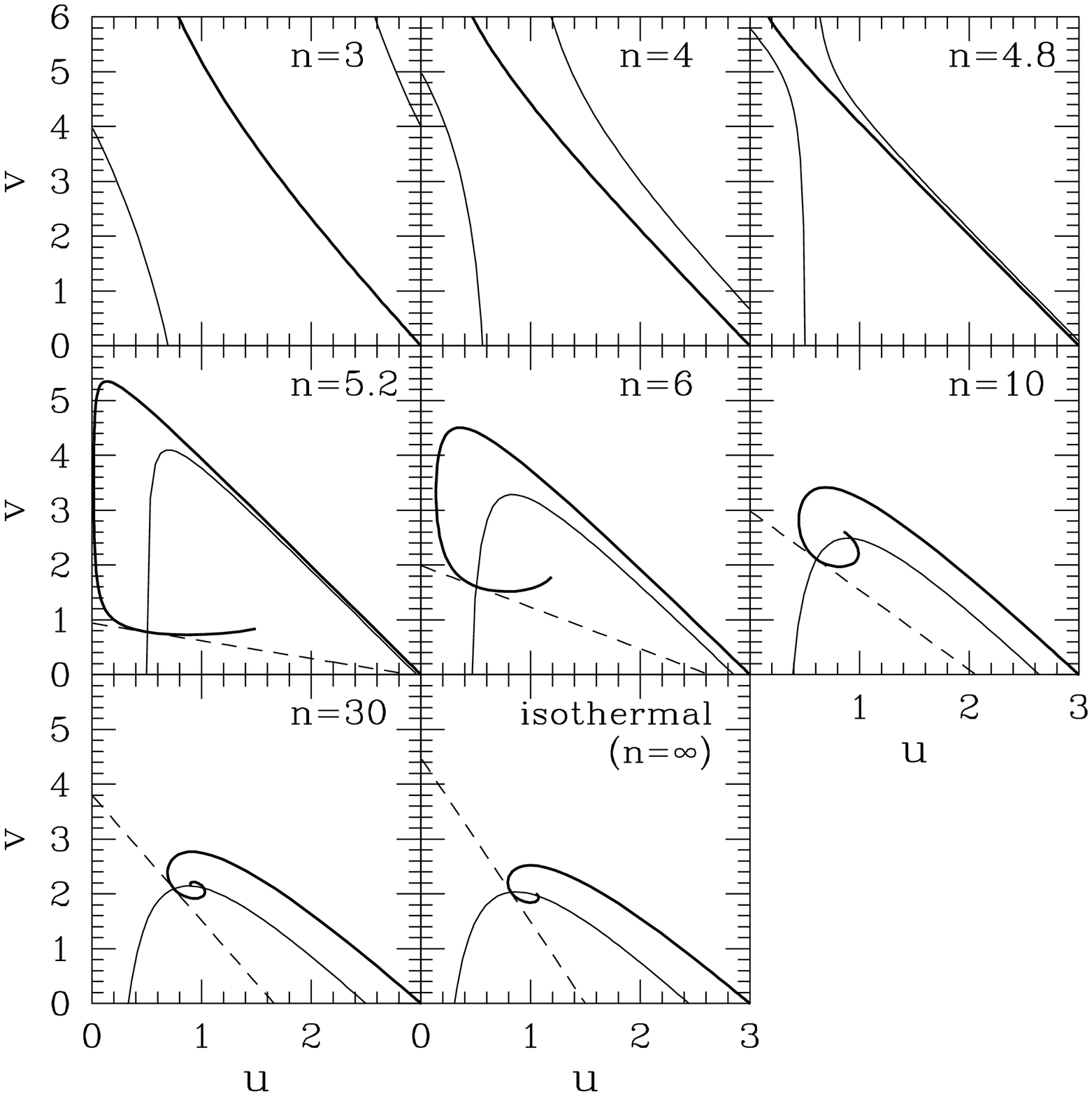}
  \end{center}
    \caption{Stability/instability criterion in $(u,v)$-plane}
    \label{fig: uv_plane2}
\end{figure}

In Figure \ref{fig: uv_plane2}, with a great advantage of
$u,v$-variables, the stability/instability criterion 
(\ref{eq: stability_!}) is examined.  
In each panel, the thick-solid lines denote 
the trajectories of the Emden solution. 
The thin-solid lines correspond to the equation $g(u,v)=0$. 
From Figure \ref{fig: uv_plane2}, one can observe that the curve
$g(u,v)=0$ drastically changes its behavior around the index $n=5$ 
(compare $n=4.8$ with $n=5.2$ case). 
Note also the different asymptotic behavior of the Emden solutions 
between the $n>5$ cases and the $n<5$ cases  
(see also Figure \ref{fig: uv_plane1}). As a consequence, 
the critical point appears when the polytrope index $n>5$ 
and it disappears in the $n<5$ cases. 

For the $n > 5$ cases, we also plot the straight line 
(\ref{eq: u-v_line}) with the 
critical value $\lambda_c$ (dashed line). 
It should be emphasized that three lines, (i.e. 
the trajectory of the Emden solution, 
the stability/instability criterion (\ref{eq: stability_!}) 
and the condition (\ref{eq: u-v_line}) with the 
critical value $\lambda_c$) intersect at 
the critical point. For comparison, the result of the isothermal gas sphere 
is plotted in Figure \ref{fig: uv_plane2}.  
The existence of the critical point still holds 
in the limit $n\to\infty$ \cite{Antonov1962,LW1968,HS1978}
\cite{Padmanabhan1989}.   
%
%
%
%
%
%
%
%
%
%
%
%
\section{Discussion and conclusions}
\label{sec: conclusion}
%
%
%
%
%
%
In this paper, we have applied Tsallis-type generalized entropy 
$S_q$ to the problem of seeking the stable distribution of 
self-gravitating systems. In contrast to previous work using the 
Boltzmann-Gibbs entropy, the local extremum state of Tsallis-type 
entropy has been found to be equivalent 
to the stellar polytrope system and Tsallis' $q$-parameter is 
related to the polytrope index $n$ (see eq.[\ref{eq: n_q}]). 
Then we apply the usual turning-point analysis to explore 
stability/instability of the local extremum state. We note that
the homology invariants $(u,v)$ are very useful to investigate 
the nature of equilibrium state of the self-gravitating system,
including the stability/instability criterion (\ref{eq: stability_!!}). 
Then we develop the second variation of entropy and re-examine the 
stability/instability of the local extremum state. The 
stability/instability criterion (\ref{eq: stability_!}),
which exactly coincides with result 
(\ref{eq: stability_!!}), can be obtained 
by solving the integro-differential equation of zero-eigenvalue state.

The results imply the important conclusion that 
the polytropic gaseous sphere 
within radius $r_e$ exhibits the gravothermal instability in the 
case of polytrope index $n>5$. The characteristic values 
$\lambda$ and $\rho_c/\rho_e$ are also evaluated at 
critical point and 
presented in Table \ref{tab: ErGM}. These values asymptotically 
approach the well-known result of the isothermal sphere in the limit
$n\to\infty$(or $q\to1$). 
In the isothermal case, the gravothermal instability can be 
interpreted in terms of a negative specific heat 
which seems to be a common feature in the 
self-gravitating systems\cite{LW1968}. We note, however,  that 
the system should have sufficient amount of outer normal part 
in order to trigger the instability; 
the specific heat for total system should be positive for 
onset of the instability, although its central part has 
negative specific heat. A similar argument holds 
for the polytropic system so that the instability appears 
for the system of $n>5$ which can have an elongated outer 
part. Evaluation of the specific heat based on 
the canonical ensemble corresponding to Tsallis' type entropy 
seems to be very important on this issue\cite{ST2001}.

Finally, we comment on the role of constraints $M$ and $E$ to the extremum 
solution (\ref{eq: df_1}). In our present analysis, standard definition of 
mean values is adopted for mass and energy 
(see eqs.[\ref{eq: mass}][\ref{eq: energy}]). 
Tsallis, Mendes and Plastino\cite{TMP1998} recently showed that this 
choice yields undesirable divergence in some physical systems including 
L\'evy random walk. To overcome the mathematical difficulty, they suggest 
that the $q$-generalized mean value should be used with a correct 
normalization instead of the standard mean value (see also \cite{MNPP2000}). 
According to their suggestion, in Ref.\cite{PP1999}, $q$-generalized, 
but {\it un-normalized} mean value was used to define $E$ and $M$. In this 
case, the resultant form of extremum state essentially remains unchanged 
and the polytropic equation of state still holds. We note, however, that 
if we use the {\it normalized} $q$-expectation value correctly, 
the problematic difficulty conversely arises. This is due to the 
fact that the potential energy $U$ associated with the distribution function 
$f(\xx,\vv)$ (see eqs.[\ref{eq: energy}][\ref{eq: potential}]) 
becomes nonlinear function of $f$, 
while the new formalism in Refs.\cite{TMP1998,MNPP2000} implicitly assumes 
the linear function. Certainly, the normalized $q$-expectation 
value will play an essential role to avoid the unexpected singular behavior, 
however, no undesirable divergence has appeared in our present analysis. 
Furthermore, it has been shown that the Tsallis formalism with standard 
linear mean values still verifies the Legendre 
transform structure, leading to the standard results of thermodynamic 
relation \cite{CT1991,PP1997}. Therefore, at least in our case of the 
self-gravitating system described by the distribution function $f$, 
the analysis using standard mean values can be validated and 
the conclusion remains correct.

The framework of nonextensive statistics based on 
Tsallis' type entropy seems to give a consistent generalization of 
the usual thermodynamical structure 
\cite{T1988,TMP1998,MNPP2000,AMPP2001,T1999}.
These works, however, are mainly concerned with construction of  
a consistent formal framework. In the light of this,  
our specific application of the Tsallis-type entropy to the 
self-gravitating systems is the first realistic physical consideration. 
Due to the spherical symmetry,
the system is simple and we can easily evaluate physical 
quantities, e.g., pressure, energy and entropy. Nevertheless, the result 
still shows a very interesting phenomenon, i.e., the gravothermal 
instability. Furthermore a fact of the gravity being a long-range force 
suggests that the self-gravitating system is one of 
the most preferable and interesting testing grounds for 
the framework of nonextensive statistics.

\section*{Acknowledgment}

The authors thank an anonymous referee for pointing out 
the crucial remark on the use of $q$-generalized mean values and 
for bringing to our attention some important references.  
We also thank Yasushi Suto for careful reading of our manuscript and 
comments.  
\clearpage
%
%
%
%
\section*{Appendix A: Some formulae for integration}
\label{appen_A0}
%
%
%
%
Here we list some formulae which have been used in section \ref{sec: poincare}
and \ref{sec: 2nd_variation}.
From the hydrostatic equations (\ref{eq: hydro_1}), 
(\ref{eq: hydro_2}) and
the integration by parts, we obtain
\begin{eqnarray}
\int_0^{r_e} dr\,\frac{Gm}{r^2}\,4\pi r^3 \rho 
&=& -\int_0^{r_e} dr\,4\pi r^3\, \frac{dP}{dr} 
= -4\pi r_e^3 P_e + 3 \int_0^{r_e}dr \,4\pi r^2\,P, 
\label{eq: int_1}
\end{eqnarray}
\begin{eqnarray}
\int_0^{r_e}  dr\,\frac{G\,m^2}{r^2} 
&=&  -\frac{GM^2}{r_e} +2 \int_0^{r_e} dr\,4\pi r^2\,\frac{Gm}{r}\rho\,
= -\frac{GM^2}{r_e} -2 \int_0^{r_e} dr\,4\pi r^3 \frac{dP}{dr}
\nonumber \\
&=& -\frac{GM^2}{r_e} \, - \,8\pi r_e^3P_e \,+ \,6 \int_0^{r_e}dr \,
4\pi r^2\,P,
\label{eq: int_2}
\end{eqnarray}
Similarly we can rewrite the integration of pressure $P$ as 
\begin{eqnarray}
\int_0^{r_e}dr \,4\pi r^2\,P 
&=& \frac{4\pi}{3}r_e^3 \,P_e -\int_0^{r_e} dr\,
\frac{4\pi}{3}\,r^3\,\frac{dP}{dr} 
= \frac{4\pi}{3}r_e^3 \,P_e + \frac{G}{6}\int_0^{r_e}
\frac{dr}{r} \,\frac{d}{dr}(m^2)
\nonumber \\
&=& \frac{4\pi}{3}r_e^3 \,P_e + \frac{GM^2}{6r_e} - \frac{n+1}{6}
\int_0^{r_e} dr \frac{d}{dr}\left(\frac{P}{\rho}\right) m
\nonumber \\
&=& \frac{4\pi}{3}r_e^3 \,P_e + \frac{GM^2}{6r_e} 
-\frac{n+1}{6} \left\{\frac{MP_e}{\rho_e} 
- \int_0^{r_e}dr \,4\pi r^2\,P \right\},
\nonumber
\end{eqnarray}
which leads to
\begin{eqnarray}
\int_0^{r_e}dr \,4\pi r^2\,P  =- \frac{1}{n-5}
\left\{8\pi\,r_e^3P_e-(n+1)\frac{MP_e}{\rho_e}+\frac{GM^2}{r_e}\right\}
\label{eq: int_3}.
\end{eqnarray}
%
%
%
%
%
\section*{Appendix B: Second variation of entropy}
\label{appen_A}
%
%
%
%
In this appendix, we derive the second variation of entropy around the
equilibrium state $\delta S_q=0$.

We first express the entropy (\ref{eq: TS_entropy}) of the extremum state. 
Substitution of the distribution function (\ref{eq: DF_Tsallis}) leads to 
\begin{eqnarray}
  S_q^{\rm(max)} &=& -\frac{1}{q-1}\int d^3\xx d^3\vv \,(f^q-f)
\nonumber\\
&=& \,\,\left(n-\frac{3}{2}\right)\,\left[M-
\frac{\tilde{B}_n}{T^{(3/2)/(n-3/2)}}\,\,\int d^3\xx\,\rho^{1+1/n}\right], 
  \label{eq: S_q_max}
\end{eqnarray}
where the constant $\tilde{B}_n$ is given by 
\begin{equation}
  \tilde{B}_n = 
\frac{B(3/2,n+1/2)}{\left\{ B(3/2,n-1/2)\right\}^{(n-1/2)/(n-3/2)}}
\,\,\left[4\sqrt{2}\pi\,\left(n-\frac{3}{2}\right)^{3/2}\right]^{-1/(n-3/2)}. 
\end{equation}

Just for convenience, we introduce the quantity $W$ 
\begin{equation}
  W= \int d^3 \xx \,\,\rho^{1+1/n}, 
\end{equation}
and vary (\ref{eq: S_q_max}) with respect to $\rho$ and $T$ keeping 
$E$ and $M$ fixed. Up to the second order, we obtain
\begin{eqnarray}
 \delta S_q &=& -\frac{\tilde{B}_n}{T^{(3/2)/(n-3/2)}}
 \,\, \left[- \frac{3}{2}\left\{\frac{W\delta T+\delta T \delta W}{T} 
 -\left(\frac{n}{n-3/2}\right)\, \frac{W}{2\,T^2}\,\,(\delta T)^2\right\}
\right.
\nonumber \\
&& \left. +\,\left(n-\frac{3}{2}\right)\,\,\delta W  \right].
\label{eq: del_Sq_max}
\end{eqnarray}
On the other hand, the constraint $\delta E=0$ gives
\begin{eqnarray}
0 = \delta E &=& \delta \left\{ 
 \frac{3}{2}\,\left(\frac{n-3/2}{n+1}\right)
T\,W\,+\,\frac{1}{2}\int d^3\xx \rho\Phi \right\} 
\nonumber \\
&=& \frac{3}{2}\left(\frac{n-3/2}{n+1}\right) \,\,
  \delta(TW) + \int d^3\xx 
  \left( \Phi\,\delta\rho +\frac{1}{2}\delta\rho\,\delta\Phi \right).
\label{eq: del_E}
\end{eqnarray}
We thus find 
\begin{equation}
  \delta T\,\, (W+\delta W) = -\frac{2}{3}\left(\frac{n+1}{n-3/2}\right)\,
  \int d^3\xx \left( \Phi\delta\rho+\frac{1}{2}\delta\rho\delta\Phi \right)
  -T \delta W.
  \label{eq: del_T}
\end{equation}
(In arriving at eq.[\ref{eq: del_E}], we have used the fact that 
$\int d^3\xx\,\,\Phi\delta\rho=\int d^3\xx\,\,\rho\delta\Phi$).
Substituting (\ref{eq: del_T}) into (\ref{eq: del_Sq_max}), we eliminate 
the quantity $\delta T$.  Then, collecting the second order terms yields the 
second variation of entropy $\delta^2S_q$ as follows: 
\begin{eqnarray}
  \delta^2 S_q &=& -\frac{\tilde{B}_n}{T^{(3/2)/(n-3/2)}}\,\,
  \left[ \frac{n+1}{n-3/2}\,\,\int d^3\xx \,\,
\frac{\delta\rho\,\delta\Phi}{2T}+ n\,\,\delta W
\right. 
\nonumber\\
&& \left.
+\frac{3}{4}\,\frac{1}{T^2\,W}\,\frac{n}{n-3/2}
 \left\{\frac{2}{3}\,\,\frac{n+1}{n-3/2}\int d^3\xx\,\,\Phi\,\delta\rho
   +T\delta W\right\}^2 \right].
\label{eq: del2_S_q}
\end{eqnarray}
Now consider the variation of $W$: 
\begin{equation}
  \delta W = \delta\left(\int d^3\xx\,\,\rho^{1+1/n} \right)\,
  =\,\frac{n+1}{n}\,\int d^3\xx\,\,\rho^{1/n}
\left(\delta\rho+\frac{1}{2n}\,\frac{(\delta \rho)^2}{\rho} \right), 
\label{eq: del_W}
\end{equation}
Neglecting the higher order contributions and keeping the second order terms 
only, substitution of (\ref{eq: del_W}) finally leads to the expression 
(\ref{eq: 2nd_S}):
\begin{eqnarray}
\delta^2S_q &=&\,\,-\frac{\tilde{B}_n}{T^{(3/2)/(n-3/2)}}\,\,
\frac{n+1}{n-3/2}\,\,
  \left[\int d^3\xx \left\{\frac{\drho\dPhi}{2T}+\frac{n-3/2}{2n}
\frac{(\drho)^2}{\rho^{1-1/n}}\right\}
\right.
\nonumber \\
&& + \left. \frac{n(n+1)}{(n-3/2)^2}\,\,
\frac{1}{3\,T^2 W} 
\left\{\int d^3\xx\left(\Phi+\frac{3}{2}\,\frac{n-3/2}{n}\,T\,\rho^{1/n}\right)
\drho\right\}^2
\right].
\end{eqnarray}

Note that the above equation indeed reduces to the well-known result 
in the isothermal sphere. Using the 
fact that $\tilde{B}_n\to1$ and $W\to M$ in the limit $n\to\infty$, 
we obtain 
\begin{eqnarray}
\delta^2S_q &\stackrel{n\to\infty}{\longrightarrow}&\,\,
  -\int d^3\xx \left\{\frac{\drho\dPhi}{2T}+
\frac{(\drho)^2}{\rho}\right\}
 - \frac{1}{3\,T^2 M} 
\left\{\int d^3\xx\Phi\drho\right\}^2, 
\nonumber
\end{eqnarray}
which is equivalent to the expression (16) of Antonov\cite{Antonov1962} and 
equation (A1.10) of Padmanabhan\cite{Padmanabhan1989}.
%
%
%
%
%
%
%
%
%
\section*{Appendix C: Positivity of minimum eigenvalue $\kappa_{\rm min}$}
\label{appen_B}
%
%
%
For the sake of completeness, in this appendix, we will prove that 
for sufficiently small $r_e$, minimum eigenvalue  
of the equation (\ref{eq: eigen_eq}), 
$\kappa_{\rm min}$ can become positive, 
irrespective of the polytrope index $n$.  

First recall that from (\ref{eq: A_quad}) and (\ref{eq: eigen_eq}), 
positivity $\kappa_{\rm min}>0$ is equivalent to the condition $A[Q]<0$. This 
gives
\begin{eqnarray}
&&-\int_0^{r_e} dr \left\{\frac{\dPhi}{2T}\frac{dQ}{dr}+
\frac{n-3/2}{n}\frac{1}{8\pi r^2
\rho^{1-1/n}}\left(\frac{dQ}{dr}\right)^2
\right\}
\nonumber \\
&&-\frac{1}{3 T^2 W} \frac{n(n+1)}{(n-3/2)^2} 
\left[ \int_0^{r_e}dr \left\{\Phi+\frac{3}{2}\left(\frac{n-3/2}{n}\right)\,T\, 
\rho^{1/n}\right\} \frac{dQ}{dr}\right]^2\,<0.
  \label{eq: A'_term}
\end{eqnarray}
In the above expression, while the second term in the right hand side of equation 
is always negative, the first term can be expressed as 
\begin{equation}
 [\mbox{1st term}] = -\frac{1}{2T}\left(H-1\right)\,\,\int_0^{r_e}
dr\,\frac{GQ^2}{r^2},  
\end{equation}
where we define 
\begin{equation}
 H \equiv \frac{\displaystyle \int_0^{r_e}dr\,\frac{1}{4\pi 
  r^2\rho}\left(\frac{P}{\rho}\right)\left(\frac{dQ}{dr}\right)^2}
  {\displaystyle \frac{n}{n+1}\,\int_0^{r_e}dr\,\frac{GQ^2}{r^2}}.
\label{eq: def_H}
\end{equation}
Thus, the inequality $H>1$ provides a sufficient condition for the
positivity of $\kappa_{\rm min}$. 

Now, we rewrite the definition (\ref{eq: def_H}) integrating by parts. 
We obtain 
\begin{equation}
 -\frac{d}{dr}\left\{\frac{1}{4\pi r^2\rho}\left(\frac{P}{\rho}\right)
\frac{dQ}{dr}\right\}  =H\,\,\frac{n}{n+1}\,\,\frac{GQ}{r^2}. 
\label{eq: eigensystem_H}
\end{equation}
The above equation can be regarded as the eigenvalue equation with 
eigenvalue $H$. In fact, one can show that equation 
(\ref{eq: eigensystem_H}) has 
minimum eigenvalue $H_{\rm min}=1$ for some small radius $r_e$. 
Consider the function 
\begin{equation}
 Q_{\rm min}=c\left\{4\pi r^3\rho(r)-\frac{n-3}{n-1}m(r)\right\}=c\,\,
  \left(u-\frac{n-3}{n-1}\right)\,m, 
  \label{eq: solution_QH}
\end{equation}
where $c$ is merely a constant. The above function 
behaves like $Q_{\rm min}\simeq c\,(2n/(n-1))\, m \to 0$ near 
the origin $r=0$ and it also becomes vanishing at $r=r_1$, where 
$u(r_1)=(n-3)/(n-1)$. Further, $Q_{\rm min}$ does not vanish 
during the interval 
$(0,r_1)$. Hence, when setting the radius of the wall 
as $r_e=r_1$, the function (\ref{eq: solution_QH}) 
indeed corresponds to the ground state of the eigenvalue equation, i.e, 
perturbation mode without any nodes. Therefore, this fact proves that 
if we suitably choose a smaller radius $r_e<r_1$, the eigenvalue of 
the system (\ref{eq: eigensystem_H}) should be larger than unity, i.e, 
$H>1$. Hence, all equilibrium configurations with $r_e<r_1$ become stable. 
\clearpage
%
%
%
%
%
%

%
%
%
%
%
%
%
%
%
%
%
%
%
%
%
%
%
%
%
%
%
%
%
%
%
%
%
\clearpage
%
%
%
%
%
%
\begin{table}
\caption{Energy-radius-mass relation and density contrast between center 
and edge evaluated at a critical point for given polytrope index $n$ or $q$}
\label{tab: ErGM}

\vspace*{0.5cm}

  \begin{center}
\begin{tabular}{|ccccc|} 
\hline
 \makebox[1.0cm]{n}  & \makebox[1.0cm]{q} &\makebox[0.1cm]{} &
 \makebox[2.5cm]{$\lambda\left(=-\frac{Er_e}{GM_e^2}\right)$} 
& \makebox[2.5cm]{$\frac{\rho_c}{\rho_e}$}\\ 
\hline\hline
5 & $\frac{9}{7}$ && ----- & ----- \\
6 & 1.22 &&  3.78 & 1.27$\times10^{6}$ \\
7 & 1.18 &&  1.69 & 1.46$\times10^{5}$ \\
8 & 1.15 &&  1.162 & 4.83$\times10^{4}$ \\
9 & 1.13 &&  0.932 & 2.40$\times10^{4}$ \\
10 & 1.12 &&  0.804 & 1.46$\times10^{4}$ \\
30 & 1.04 &&  0.429 & 1590 \\
50 & 1.02 &&  0.388 & 1130 \\
100 & 1.01 &&  0.360 & 887 \\
$\infty$ & 1 &&  0.335 & 709 
\\ \hline
\end{tabular}
  \end{center}
\end{table}
%
%
%
%
%
%
%
\clearpage
%
%
%
%
%
%
%
%
%
%
%
%
%
%
%
%
%
%

\begin{thebibliography}{00}
%
%
%
\bibitem{Antonov1962} V.A. Antonov, {\it Vest. Leningrad Gros. Univ.}, 
7 (1962) 135 (English transl. in {\it IAU Symposium 113, Dynamics of 
Globular Clusters}, ed. J. Goodman and P. Hut [Dordrecht: Reidel], 
pp. 525--540 [1985]) 
\bibitem{LW1968} D. Lynden-Bell, R. Wood, Mon.Not.R.Astr.Soc. 138 (1968) 495.
\bibitem{HS1978} I. Hachisu, D. Sugimoto, Prog.Theor.Phys. 60 (1978)
        123.  
\bibitem{HNNS1978} I. Hachisu, Y. Nakada, K. Nomoto, D. Sugimoto,
        Prog.Theor.Phys. 60 (1978) 393.  
\bibitem{Padmanabhan1989} T. Padmanabhan, Astrophys.J.Suppl. 71 (1989)
        651. 
\bibitem{Padmanabhan1990} T. Padmanabhan, Phys.Rep. 188 (1990) 285.
        651. 
\bibitem{BS1984} E. Bettwieser, D. Sugimoto, Mon.Not.R.Astr.Soc. 208
        (1984) 493.
\bibitem{BT1987} J. Binney, S. Tremaine, {\it Galactic Dynamics} 
(Princeton Univ. Press, Princeton, 1987) 
\bibitem{EPI1987} R. Elson, P. Hut and S. Inagaki, Ann. Rev. Astron. Astrophys. 25 (1987) 565.
\bibitem{MH1997} G. Meylan, D.C. Heggie, Astron.Astrophys.Rev. 8 (1997) 1.
\bibitem{HD1994} X.P. Huang, C.F. Driscoll, Phys.Rev.Lett. 72 (1994) 2187. 
\bibitem{SZF1995} M.F. Shlesinger, G.M. Zaslavsky, U.Frisch, {\it Levy 
    Flights and Related Topics in Physics} (Springer, Berlin, 1995)
\bibitem{K1997} I. Koponen, Phys.Rev.E 55 (1997) 7759.
\bibitem{T1988} C. Tsallis, J.Stat.Phys. 52 (1988) 479.
\bibitem{CT1991} E.M.F. Curado, C. Tsallis, J.Phys.A 24 (1991) L69. 
\bibitem{PP1997} A. Plastino, A.R. Plastino, Phys.Lett. A 226 (1997) 257.
\bibitem{TMP1998} C. Tsallis, R.S. Mendes, A.R. Plastino, 
Physica A 261 (1998) 534.
\bibitem{MNPP2000} S. Mart\'inez, F. Nicol\'as, F. Pennini, A. Plastino, 
Physica A 286 (2000) 489.
\bibitem{AMPP2001} S. Abe, S. Mart\'inez, F. Pennini, A. Plastino, 
Phys.Lett. A 281 (2001) 126.
\bibitem{B1996} B.M. Boghosian, Phys.Rev. E 53 (1996) 4754.
\bibitem{LT1998} M.L. Lyra, C. Tsallis, Phys.Rev.Lett. 80 (1998) 53.
\bibitem{AA2001} T. Arimitsu and N. Arimitsu, Physica A 295 (2001) 177.
\bibitem{T1999} C. Tsallis, Braz. J. Phys. 29 (1999) 1.
\bibitem{AO2001} S. Abe, Y. Okamoto (Eds.), {\it Nonextensive Statistical Mechanics and Its Applications} (Springer, Berlin, 2001) 
\bibitem{HH2001} C. Hanyu, A. Habe, Astrophys. J. 554 (2001) 1268.
\bibitem{LKRQT1998} A. Lavagno, G. Kaniadakis, M. Rego-Monteiro, P. Quarati, C. Tsallis, Astrophys.Lett.Comm. 35 (1998) 449. 
\bibitem{PP1993} A.R. Plastino, A. Plastino, Phys.Lett. A 174 (1993) 384. 
\bibitem{PP1999} A.R. Plastino, A. Plastino, Braz. J. Phys. 29 (1999)
        79.  
\bibitem{Chandra1939} S. Chandrasekhar, {\it Introduction to the Study of
        Stellar Structure} (New York, Dover, 1939) 
\bibitem{KW1990} R. Kippenhahn, A. Weigert, {\it Stellar Structure 
and Evolution} (Springer, Berlin, 1990)
\bibitem{ST2001} M. Sakagami and A. Taruya, in preparation.
\end{thebibliography}
\end{document}